\documentclass[useAMS,usenatbib]{mnras} 
\usepackage{amsmath}
\usepackage{graphicx}
\usepackage[T1]{fontenc}
\usepackage{epstopdf}
\graphicspath{{../Figures/}}

\title[]{The quenching and survival of ultra-diffuse galaxies in the Coma cluster}
\author[C. Yozin and K. Bekki]{C. Yozin\thanks{{\bf E-mail} 21101348@student.uwa.edu.au; kenji.bekki@uwa.edu.au} and K. Bekki\footnotemark[1]
\\\\
ICRAR, M468, The University of Western Australia, 
35 Stirling Highway, Crawley
Western Australia, 6009, Australia}

\begin{document}
\date{Accepted 2015 January. Received 2015 January; in original form 2015 January}
\pagerange{\pageref{firstpage}--\pageref{lastpage}} \pubyear{2013}
\maketitle
\label{firstpage}

\newcommand{\bia}{$b/a$}
\newcommand{\cnfw}{c$_{\rm NFW}$}
\newcommand{\ex}[1]{10$^{\rm #1}$}
\newcommand{\ih}{$h^{\rm -1}$}
\newcommand{\kms}{kms$^{\rm -1}$}
\newcommand{\lcdm}{$\Lambda$CDM}
\newcommand{\lsol}{L$^*$}
\newcommand{\maga}{mag arcsec$^{\rm -2}$}
\newcommand{\mst}{M$_{\rm s}$}
\newcommand{\mex}[2]{#1$\times$10$^{\rm #2}$}
\newcommand{\msol}{M$_{\odot}$}
\newcommand{\muc}{$\mu$(g,0)}
\newcommand{\mur}{$\mu$(g,r)}
\newcommand{\mvir}{M$_{\rm vir}$}
\newcommand{\pcsq}{pc$^{\rm -2}$}
\newcommand{\reff}{r$_{\rm e}$}
\newcommand{\rs}{r$_{\rm s}$}
\newcommand{\rvir}{r$_{\rm vir}$}
\newcommand{\vrot}{V$_{\rm rot}$}
\newcommand{\z}[1]{$z=#1$}


\begin{abstract}

We conduct the first self-consistent numerical simulations of a recently discovered population of 47 large, faint (ultra-diffuse) galaxies, speculated to lie in the Coma cluster. With structural properties consistent with very large low surface brightness systems (i.e. \muc{}$<24$ mag arcsec$^{\rm -2}$; \reff{} comparable to the Galaxy), the red colour ($\langle$g-r$\rangle$$\sim$0.8) and assumed low metallicity of these objects compels us to consider a scenario in which these are underdeveloped galaxies whose early ($z$$\simeq$2) accretion to an overdense environment quenched further growth. Our simulations demonstrate the efficacy of this scenario, with respect to available observational constraints, using progenitor galaxy models derived from scaling relations, and idealised tidal/hydrodynamical models of the Coma cluster. The apparent ubiquity of these objects in Coma implies they constitute an important galaxy population; we accordingly discuss their properties with respect to a \lcdm{} cosmology, classical LSBs, and the role of baryonic physics in their early formation.

\end{abstract}


\section{Introduction}

Low surface brightness galaxies (LSBs) remain an important test for the favoured \lcdm{} cosmology, given the dominance of dark matter in these systems whose number density exceeds that of normal galaxies \citep{both97, dalc97b}. Although distinguished by classification from high-SB systems and exhibiting qualitatively different halo properties at low radii \citep[e.g.][]{debl01}, their conformance to the Tully-Fisher relation  supports a continuity with late-type discs \citep{zwaa95, scho14}. In the standard galaxy formation framework, these diffuse systems nominally formed within low initial density fluctuations \citep{fall80, moma98}, resulting in blue, gas-rich and slowly evolving discs \citep{mcga94b}. The vulnerability of such tenuous systems to external influence typically precludes their existence in high-density environments \citep{deke86, rose09, gala11}.

Improved imaging techniques reveal however a surprising new population of such galaxies within rich cluster environments. Specifically, we refer to the recent detection in the Dragonfly Telephoto array of large diffuse galaxies \citep[][VD15]{vdok2015a} in the Coma cluster, whose association with the cluster is proposed on the basis of both their spatial distribution and a lower limit on their distance as implied by the unresolved nature of their stellar component. With further examination via high quality CFHT imaging, and assuming a distance equivalent to Coma, the authors assert 47 of these objects as Ultra-diffuse galaxies (UDGs) with effective radii (\reff{}) ranging from 1.5 to 4.5 kpc, and central surface brightnesses of 24 to 26 mag arcsec$^{\rm -2}$ (Table 1). The membership of the largest of these objects in Coma has since been confirmed using spectroscopy from the Keck I telescope \citep{vdok2015b}. 

While classical blue LSBs have been previously detected in the Virgo cluster \citep{impe88}, the UDGs lie on the faint end of the red sequence of the (more massive) Coma cluster \citep{gava13}, a dichotomy possibly related to environmental dependencies. The recent accretion and quenching of gas-rich LSBs in Coma is consistent with 1) the slow formation of LSBs within cosmologically underdense regions; 2) their recent accretion to large-scale structures/filaments \citep{rose09}, and 3) a recent build-up of the faint end/an increase in the dwarf-to-giant ratio in the cluster red sequence since \z{0.2}{} \citep{lu09, gava13}. Moreover, other faint dwarves in Coma display generally radial/anisotropic orbits, suggestive of recent accretion from the field or as part of subgroups \citep{adam09}.

If applying stellar population models, however, the median UDG colour $\langle$g-r$\rangle=0.8\pm0.1$ is consistent with an old stellar disc passively evolving over a timescale up to \z{2}{}, if assuming the typically low metallicity of diffuse LSBs \citep[i.e. oxygen abundance 12+log(O/H)$\leq$8;][]{mcga94b}. The paucity of quenched field galaxies within this mass range \citep{geha12} thus implies a scenario in which the UDGs were quenched upon infall at high redshift, leading VD15 to speculate that these are examples of failed L$_*$-type galaxies. 

This early quenching is supported by evidence of an almost complete red sequence in a \z{1.8}{} analogue of Coma \citep[JKCS041;][]{andr14}. The apparent absence of UDGs within 300 kpc of the cluster centre, presumably via their destruction here, is also not consistent with their very recent accretion to a kinematically hot host. More generally, the apparent ubiquity of these UDGs in Coma conflicts with the theoretical rarity of large diffuse LSBs such as Malin I \citep{hoff92}, motivating an alternate scenario for their origin.

In this first theoretical study of these objects, we adopt self-consistent numerical methods for hypothetical UDG progenitors to ascertain their evolution within high-density environments modelled on the Coma cluster. Guided by observational constraints from VD15, we devise a theoretical template for the scenario in which these objects are normal galaxies accreted as satellites and quenched soon after the standard epoch of disc formation (\z{2}). Section 2 describes our UDG model, while Sections 3 and 4 describe its hydrodynamical and tidal interactions within a rich cluster. Section 5 concludes this study with a discussion of these results.

\section{A hypothetical model for a UDG progenitor}

\begin{table} 
\centering
\caption{Summary of a) properties of observed UDGs b) our reference model parameters}
\begin{tabular}{@{}lr@{}}
\hline
(a) UDG Parameters & Mean (Range) \\
\hline
\muc/mag arcsec$^{\rm -2}$ & 25 (24-26)   \\
\reff{}/kpc & 3 (1.5-4.5)  \\
S\'{e}rsic index, $n$ & 1 (0.5-1.5) \\
Stellar mass, \mst{}/\msol{} & \mex{6}{7}{} (\mex{1}{7}-\mex{3}{8}) \\
Cluster-centric radius/kpc & - ($>$300)  \\
\hline
(b) UDG Model Parameters & Value \\
\hline
N$^{\rm o}$. DM halo particles & \mex{1}{6}{} \\
N$^{\rm o}$. Stellar particles &  \mex{3}{5}{} \\
N$^{\rm o}$. Gas particles & \mex{2}{5}{} \\
DM halo mass, M$_{\rm h}$ (\msol{}) & \mex{3.2}{10}{} \\
NFW concentration, \cnfw{} & 5.0 \\
DM halo \rvir{} (kpc) &  83 \\
Stellar (disc) mass, M$_{\rm d}$ (\msol) & \ex{8}{} \\
Stellar scalelength, r$_{\rm d}$ (kpc) & 1.7 \\
Stellar scaleheight, z$_{\rm d}$ (kpc) & 0.34 \\
Gas mass, M$_{\rm g}$ (\msol) & \mex{5}{8}{} \\
Gas scalelength, r$_{\rm g}$ (kpc) & 5.1 \\
Gas scaleheight, z$_{\rm g}$ (kpc) & 0.34 \\
\hline
\end{tabular}
\end{table} 

\begin{figure}
\includegraphics[width=1.\columnwidth]{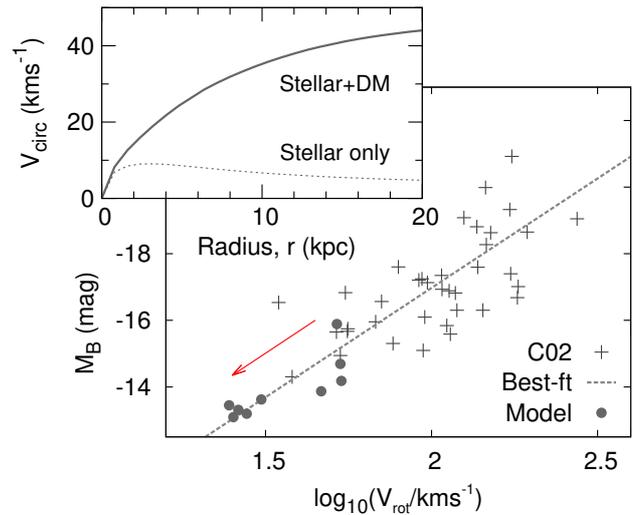}
\caption{(Top) Initial rotation curve (assuming $V(r)=(GM(<r)/r)^{\rm 0.5}$) for the UDG progenitor model; (bottom) simulation BOUND in the \vrot{}/$B$-band Magnitude phase space (black filled circles); its evolution (with direction highlighted with the red arrow) is consistent with an observed LSB sample (black crosses) with best-fit curve (dashed line), reproduced from \citet{chun02}.}
\end{figure}

Our study adopts a numerical method introduced in \citet{bekk11} and \citet{bekk14}. The N-body idealisation of the UDG progenitor is constrained by the observed properties (Table 1), in particular the estimated median stellar mass of \mex{6}{7}{} \msol{}, inferred from their median $\langle g-r \rangle$ and a tight mass-colour relation revealed in the GAMA survey. 

For an initial stellar (disc) mass (M$_{\rm d}$) of \ex{8}{} \msol{}, abundance matching in cosmological simulations \citep[e.g.][]{muns13, behr13} suggest a dark matter (DM) halo mass (M$_{\rm h}$, within the virial radius \rvir{}) up to $\sim$\ex{3}{}M$_{\rm d}$, although rotation curves obtained from observed samples suggest these simulations produce too much substructure, attaining instead a factor 10 lower prediction \citep{mill14}. We adopt an intermediate value for our UDG progenitor, and assume a mean density comparable to the Galaxy \citep[assuming r$_{\rm vir,MW}=258$ kpc and M$_{\rm h,MW}=1\times$10$^{\rm 12}$ \msol{}][]{klyp02}, giving a \rvir{} of 83 kpc. 
We assume an exponential disc morphology, with a density defined for the cylindrical radius $r$ and height $z$:
\[
\rho(r,z)\propto\exp\left(-\frac{r}{r_{\rm s}}\right){\rm sech}^2\left(\frac{z}{0.2r_{\rm s}}\right).
\]
The scalelength r$_{\rm s}$ is derived a size-mass$^{\alpha}$ scaling relation \citep{dutt09, ichi12} whose shallow slope at this stellar mass range ($\alpha\simeq0.15$, with respect to simple self-similar models based on virial relations where $\alpha\simeq0.33$) is believed to reflect the role of feedback \citep{moma98, deke03, shen03}. 

Assuming r$_{\rm eff}=1.67$r$_{\rm d}$, the sizes of the observed UDGs can be attained if we thereafter apply a linear scaling ($\lambda_{\rm s}$) to r$_{\rm d}$. If the UDGs are assumed as classical LSBs, $\lambda_{\rm s}$ is a proxy for variations in the initial spin parameter; our adopted $\lambda_{\rm s}=1.5$ lies at $\sim$1$\sigma$ of typical spin distribution among simulated halos \citep{macc07}, and is consistent with the tendency for high spin halos to host LSB galaxies. 

This selection of free parameters (M$_{\rm d}$, $\lambda_{\rm s}$) yield a central surface brightness consistent with those observed (24$<$$\mu$(B,0)/(mag arcsec$^{\rm -2}$)$<$26). We further note that the resulting ratio \reff/\rvir{} lies within 1$\sigma$ of a tight linear relation identified by \citet{krav13}, consistent with the galaxy size being set by the halo's initial specific angular momentum \citep{moma98}. 

We adopt the NFW density profile for the halo, defined as:
\[
\rho(r)\propto(r/r_{\rm h})^{\rm -1}(1+(r/r_{\rm h})^{\rm 2})^{\rm -1},
\]
where r$_{\rm h}=r_{\rm vir}$/\cnfw{} and \cnfw{} is a concentration factor, here set to 5 in accordance with mass-redshift-dependent relations for the adopted M$_{\rm d}$ \citep[i.e.][]{macc07, muno11}. This is similar to the characteristic c$_{\rm NFW}$ among a sample of LSBs obtained from fits to high quality H$_{\alpha}$-H$_{\sc I}$ \citep[${\rm c}_{\rm NFW}\simeq5-6$;][]{mcga03}, although the authors argue that a cored pseudo-isothermal density profile provides a better fit than NFW. 

The interstellar medium (ISM) is considered isothermal (temperature 10$^{\rm 4}$ K), and modelled with smoothed particle hydrodynamics. We assume again an exponential density profile, as for the disc, and adopt a scalelength r$_{\rm g}=2.6$r$_{\rm d}$  \citep[i.e. the sample mean from][]{krav13}, with a total mass (M$_{\rm g}$) five times that of the disc, as inferred from scaling relations identified by \citet{popp14}. 

Figure 1 illustrates the initial rotation curve of our model (with parameters summarised in Table 1). We do not readily find other curves of comparable extension and \vrot{} in the literature, noting also that, similar to previous simulations of a similar nature \citep[e.g.][]{kaza11}, \vrot{} will decline significantly due to substantial mass stripping. The curve is qualitatively similar to LSBs, in which the stellar disc is dynamically insignificant \citep{both97}. Our model also lies on the same Tully-Fisher relation exhibited in the sample of LSBs compliled by \citet{chun02}.

Following the method of \citet{yozi14}, our SF/feedback model \citep[see][for more detail]{bekk14} is selected principally to avoid gas clumping in the gas-rich ISM while providing the low SF rates of generic LSBs \citep{scho13}. To summarise, star formation occurs in the event of a convergent, cool region of gas with local density threshold $>$\ex{0.5}{} cm$^{\rm -3}$, while the thermal component of Supernova feedback (\mex{0.9}{51}{} ergs) is injected into the ISM over an adiabatic expansion timescale of \ex{6}{} yrs.

\section{Ram pressure stripping in a Coma-analogue cluster}

\begin{figure}
\includegraphics[width=1.\columnwidth]{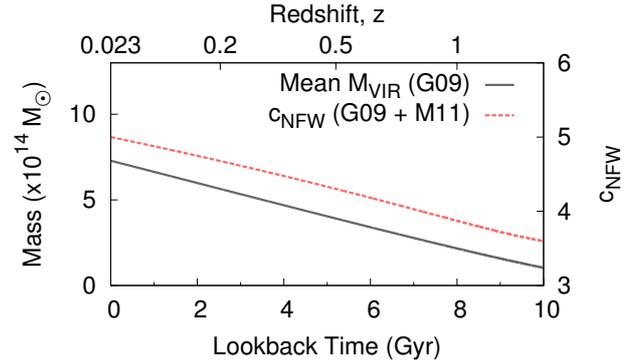}
\caption{Coma virial mass and concentration, \cnfw{}, as a function of lookback time/$z$, as derived from weak lensing methods \citep{gava09} and cosmological simulations \citep{muno11}.}
\label{f1}
\end{figure}

\begin{figure}
\includegraphics[width=1.\columnwidth]{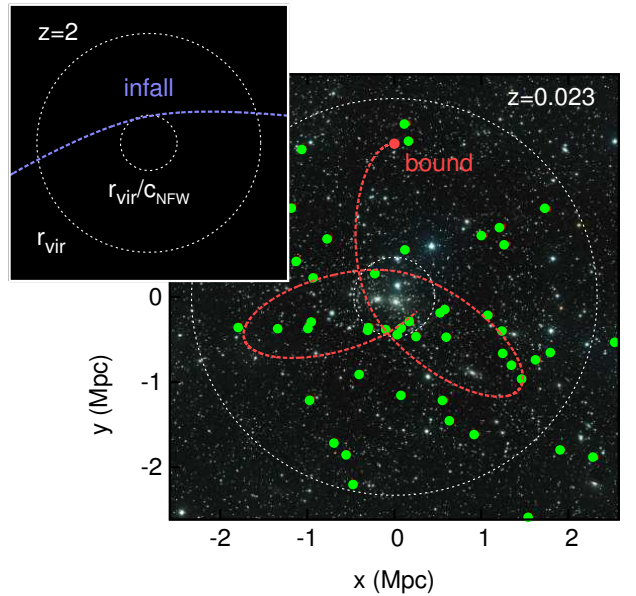}
\caption{Equal-scale schematics of Coma-cluster models considered in this study, at \z{0.023}{} and \z{2}{} (10 Gyr ago). White dotted lines represent characteristic NFW and virial radii. BOUND and infall orbits are conveyed with dashed lines, overlaid in the \z{0.023}{} case, on an optical image of Coma with green dots showing the locations of observed UDGs \citep[reproduced from][]{vdok2015a}.}
\label{f1}
\end{figure}

In this section, we test the hypothesis that the UDG progenitor model can be efficiently quenched by ram pressure stripping (RPS) upon first infall and interaction with the intracluster medium (ICM) of a predecesor of Coma at \z{2}{}. We derive a model of Coma using recent weak lensing measurements from deep CFHT images, which reveal a dynamical mass M$_{\rm coma}$ (at \z{0.024}{}) of \mex{5.1}{14}{} h$^{\rm -1}$\msol{} \citep{gava09}. Assuming H$_{\rm 0}=70$ \kms{} Mpc$^{\rm -1}$, this estimate is low compared to previous studies \citep[i.e.][]{kubo07} and is thus conservative within the context of our work. We adopt statistical measures of halo mergers/assembly, followed in cosmological simulations \citep{fakh10, muno11, ludl13} to attain the evolution of mass and \cnfw{} as a function of redshift, with respect to their present values (Figure 2). 

To model the impact of hydrodynamic interactions on our model, we adopt the method of \citet{bekk14}, in which a cluster satellite lies within a cubic lattice (with periodic boundary conditions) representing its local ICM. The SPH particles comprising this lattice have their density (and velocity, relative to the fixed satellite) vary according to a cluster-centric radial dependence described by a NFW profile (with \cnfw{} as inferred above). For the extrapolated M$_{\rm coma}$ at \z{2}{}, we obtain a total baryonic mass M$_{\rm ICM}$ from \citet{lin12}, who use X-ray emission to establish 0.1$<$M$_{\rm ICM}$/M$_{\rm coma}$$<$0.15, and an ICM kinetic temperature interpolated from the cluster sample of \citet{mats00}.

We assume the UDG progenitor is accreted onto the cluster with the cosmologically most common trajectory upon first infall \citep{bens05}. Described in terms of the radial/tangential velocity at \rvir{}, this trajectory does not vary significantly since \z{2}{}. To avoid setting our satellite in thermal equilibrium with the ICM, the orbit commences at 1.5\rvir{}, and is shown schematically in Figure 3.

We find this scenario satisfies the hypothesis that rapid quenching of the UDG progenitor can occur at high $z$; their red colour can be facilitated by the drop in the SF rate upon first infall (relative to an isolated counterpart, as illustrated in Figure 4) and consistent with other numerical studies \citep[e.g.][]{cen14}. This occurs in spite of the exclusion, in this simulation, of cluster tides that would otherwise diminish the halo restoring force. While not all gas exceeds escape speed, being retained in the massive halo, its gas surface density is insufficient to permit further SF (generally lying below the typical threshold density for SF of $\sim$10 \msol pc$^{\rm -2}$). The RP stripped ISM shows a qualitatively similar (jellyfish) morphology to Virgo cluster satellites undergoing a similar process \citep[e.g.][]{chun07}. 

The uncertainty introduced by the extrapolation of Coma's properties at higher $z$ should be highlighted, given for example, the uncharacteristically high ICM temperature of Coma for its X-ray luminosity \citep{pimb14}. Similarly, a recent analysis of the GIMIC cosmological simulations for quenched galaxies residing in (10$^{13-15}$ \msol{}) host halos illustrates a highly variable RP efficiency with redshift \citep[with primary cause as yet undetermined;][]{bahe15}. The authors note, however, that quenched galaxies preferentially lie in overdense regions of the ICM \citep[by a factor 10, with respect to the mean at a given cluster-centric radius; see also][]{tonn08} in which case our model, devoid of such substructure, can be deemed conservative.

We further emphasise that our model is of similar mass to the dwarf-type model of \citet{bekk14}, which was also demonstrated to be efficiently quenched in a Coma-type cluster environment during first infall. That his dwarf-type model was HSB, as contrasted with our high spin/diffuse model, would further indicate that the quenching efficiency demonstrated here is not sensitive to our assumptions regarding the construction of the UDG progenitor model.

\begin{figure}
\includegraphics[width=1.\columnwidth]{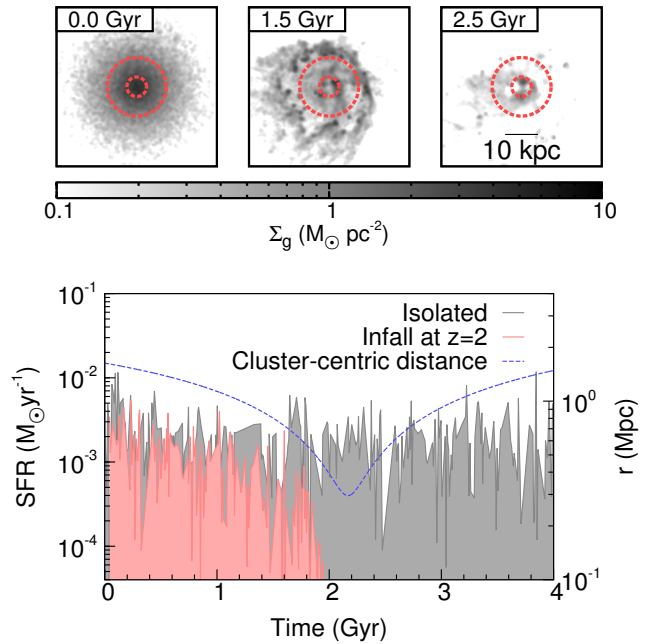}
\caption{(Top panels) Face-on gas surface mass density distribution for the UDG progenitor during first infall into a Coma-analogue at \z{2}{}. The simulation time is denoted at the top left, with effective and optical radii shown with red dotted lines; (bottom) star formation history of the aforementioned simulation (red, with cluster-centric radius shown with the blue dashed line), compared against an isolated example (grey).
}
\label{f1}
\end{figure}

\section{Survival of UDG progenitors in cluster tidal environments}

\begin{figure}
\includegraphics[width=1.\columnwidth]{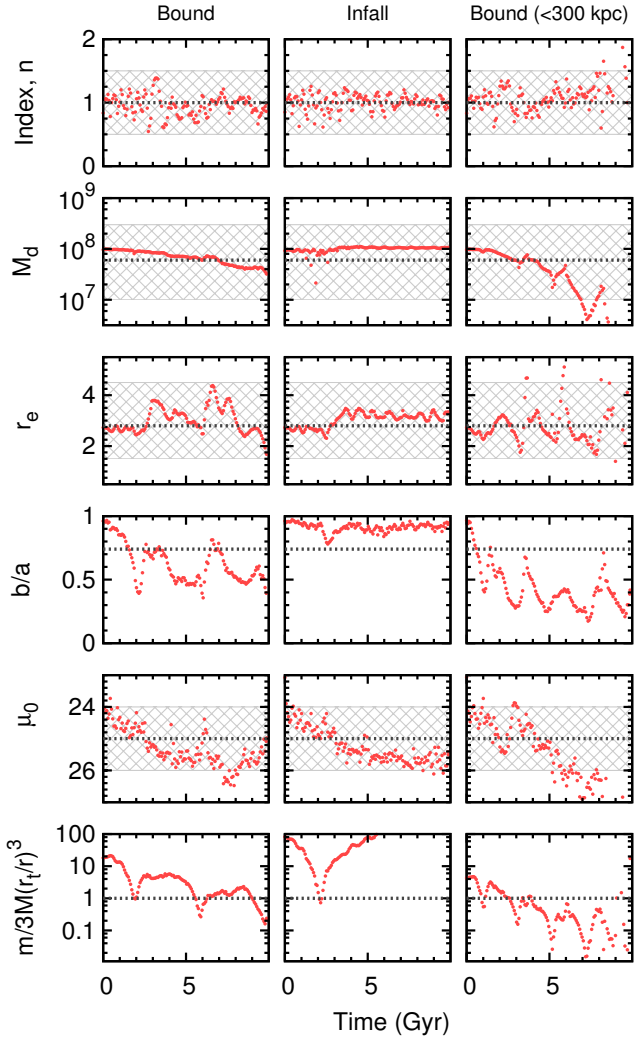}
\caption{A comparison of parameters from simulations (left to right) 'bound', 'infall', and 'bound ($<$300 kpc)', against observational constraints (Table 1). From top to bottom: S\'{e}rsic index $n$, stellar mass within g-band limit M$_{\rm d}$, effective radius r$_{\rm e}$, axial ratio b/a, $g$-band central surface brightness \muc{}, and the ratio of the mass (within r$_{\rm t}$) to the mass required to avoid tidal pruning.}
\label{f1}
\end{figure}

Our UDG progenitor is hypothesised to reside in Coma over several crossing timescales ($\tau_{\rm cr}\simeq$1 Mpc$/$1000 kms$^{\rm -1}$$\simeq$1 Gyr). Previous studies have highlighted how successive pericentre passages or harassment by other satellites can transform (S\'{e}rsic index) $n=1$ discs to earlier-type morphologies \citep{moor96, kaza11}. In this section, we compare the time-evolution of our models structural properties (illustrated in the diagnostic of Figure 5) with observed UDG constraints (Table 1). 

The group's tidal influence is modelled explicitly at each simulation time-step with the acceleration imposed upon the constituent stellar/halo particles of the UDG progenitor by the cluster halo's gravitational potential. We adopt the fixed spherically-symmetric NFW profile of Section 3 for the halo mass distribution, and derive the gravitational potential at the location of an individual progenitor particle with its cluster-centric radius. 

The trajectory of the progenitor within the cluster potential is defined by a selection of orbits/redshifts. Orbit 'infall' refers again to the cosmologically most common infall trajectory at \z{2}{}, adopted in Section 3 \citep{bens05}. Orbit 'bound' adopts the \z{0.023}{} properties of Coma, with a ratio of apocentre to pericentre (r$_{\rm peri}$) set to the median value determined from large-scale simulations \citep[6; e.g.][]{ghig98}, and r$_{\rm peri}$ fixed at 300 kpc, similar to the apparent limit identified by VD15. We accordingly test if UDG progenitors are destroyed below this limit with orbit 'bound ($<$300 kpc)', in which r$_{\rm peri}$ is 150 kpc. We assume that the UDG model would occupy these bound orbits subsequent to its quenching during the initial parabolic orbit 'infall', as a result of dynamical friction experienced through interaction with a live cluster halo (not modelled here).

Figure 5 conveys how, for our initial stellar disc mass M$_{\rm d}={\rm 10}^{\rm 8}$ \msol{}, the models following orbits 'bound' and 'infall', remain within observed ranges for UDGs. This is in spite of tidal mass loss (estimated from the stellar mass exceeding escape speed) and the passive fading of the stellar disc. We estimate the latter from stellar evolution models \citep{bruz03, port04}, assuming an initial stellar metallicity of 0.03 Z$_{\odot}$, comparable to faint Local Group satellites at \z{2}{} \citep{leam13} and consistent with passive fading since \z{1-2}{} to provide the observed ($\langle$g-r$\rangle$) colour (VD15).

These models remain disc-like and extended, consistent with VD15 who obtain a good fits in GALFIT for stacked $\Sigma_g$/$\Sigma_i$ images of the UDGs. These, together with deeper imaging and a multicomponent fit of the largest UDG \citep{vdok2015b}, exhibit low $n$ in the range 0.5 to 1.5. Unlike in our similar study, in which \ex{9}{} \msol{} dwarf satellites develop strong asymmetries and bulges within a group environment (Yozin \& Bekki, in prep.), the disc stabilisation provided by a dominant halo, together with early removal of the ISM, leads to a persistently late-type morphology. These models also tightly match the Tully-Fisher relation, established from a sample of LSBs by \citet{chun02}, for the duration of their evolution (Figure 1).

The deviations from a unity $n$ are caused principally by elongation of the disc and/or formation of tidal arms. These features are quantified with an axis ratio (b/a) estimated with the $\mu$-weighted mean axial ratio among ellipses fit to isophotes (between the satellites' central $\mu$ and $\mu_{\rm g}=28$ mag arcsec$^{\rm -2}$ at 0.25 mag arcsec$^{\rm -2}$ increments) of the smoothed face-on $\mu$ distribution (using a gaussian kernel that matches the $\sim$0.12 kpc FHWM of the CFHT imaging). The simple morphology of the models (lacking strong bisymmetries like bars due to a low disc mass) means this simple method is sufficiently robust for our purposes, as illustrated in Figure 6. For orbit 'bound', which we deem conservative in light of successive r$_{\rm peri}$ at 300 kpc, b/a falls to the lower end permitted by observations (Figure 7); by contrast, 'infall' lies at the upper end. This metric alone therefore suggests the UDGs occupy orbits in between these extremes, although we do not account for the effects of their inclination.

For orbit 'bound ($<$300 kpc)', we can demonstrate the effective destruction of the UDG progenitor, denoted by substantial mass loss and disruption to the previously exponential disc, clearly illustrated in mock CFHT images that do not compare well with the observed counterparts (Figure 6). The systems do not, however, resemble the ultra-compact dwarves often found concentrated at their respective cluster centres \citep{drin03}. Although occupying a similar magnitude range to the UDGs, the small sizes of UCDs ($<$100 pc) are speculated to reflect the loss of up to 98 percent of their original luminosity via tidal stripping \citep{bekk01}. 

The bottom row of Figure 5 shows how this satellite disruption is related to an insufficient mass (m$_{\rm t}$) enclosed within its tidal radius r$_{\rm t}$ when interacting with the cluster tidal field at pericentre. For an orbital radius $r$, tides exerted by the cluster mass $M$ enclosed within $r$ can yield substantial pruning of the satellite if m$_{\rm t}$/(3M(r$_{\rm t}$/$r$)$^{\rm 3}$)$\leq$1 \citep{binn08}. For orbits 'bound' and 'infall', we find this mass ratio of the order of unity for several $\tau_{\rm cr}$, consistent with their apparent robustness in morphology. Incidentally, we note that VD15 adopt this same formulation to predict, for the UDGs estimated stellar mass, a stellar-halo mass ratio (M$_{\rm d}$/M$_{\rm h}$) to account for these objects survival at 300 kpc; their value of $\sim$0.04 within r$_{\rm t}$ is consistent with the initial conditions of our progenitor models (Section 2).

\begin{figure}
\includegraphics[width=1.\columnwidth]{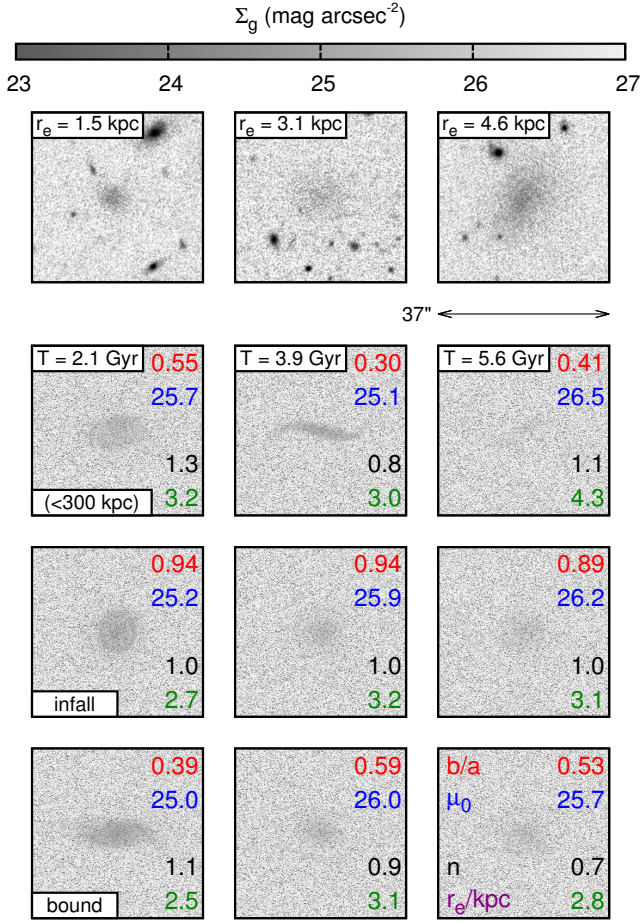}
\caption{(Top) CFHT imaging of UDGs, illustrating the range in size \citep[reproduced from][]{vdok2015a}; (bottom panels)
mock CFHT images ($g$-band surface brightness) of (from bottom to top) 'bound', 'infall', and 'bound ($<$300 kpc)', viewed face-on at simulation time ($T=$) 2.1 Gyr (first pericentre passage), 3.9, and 5.6 Gyr (second passage
). Coloured labels provide the instantaneous \bia{}, \muc{}, $n$ and \reff{}.}
\label{f1}
\end{figure}

\section{Discussion and conclusions} 

Using constraints from the first detections of large faint galaxies (UDGs) in the Coma cluster (VD15), we have developed a theoretical template for their evolutionary history. The exceptionally diffuse stellar component would imply a late infall as deemed common theoretically and observationally among LSBs \citep{rose09}, yet the degeneracy associated with their red colour permits a passive fading since as early as \z{1-2}{} (depending on the assumed metallicity). In this context, we have examined the intriguing possibility that these UDGs are an example of satelite-quenching at high redshift.

While there are no previous examples of large red LSBs in clusters, the UDGs possess several characteristics ($n$, $\mu$(g,0), b/a) similar to Local Group dSphs (VD15). It is intriguing therefore that in an inventory of local resolved stellar populations, the Galactic companions typically formed 30-50 percent of their mass (but in some cases, only 10-20 percent) by \z{2} \citep{weis11}. This downsizing (lower mass galaxies showing more prolonged formation epochs) supports the role of reionisation in these objects' early formation, where gas heating by the UV-background suppresses SF by both preventing gas collapse/accretion and heating the ISM, delaying cooling until \z{1}{} \citep{babu92, skil03}. The recent finding of a central depression in the surface brightness profile (relative to an exponential disc) of the largest UDG by \citet{vdok2015b}, a feature common amongst dSphs, leads them to speculate further the role of early stellar feedback in supressing SF at high-$z$ and contributing to gas loss due to ram pressure stripping \citep[e.g.][]{stin13}.

It remains to be demonstrated if these mechanisms would be efficient for our UDG progenitor (with V$_{\rm max}$ of 40-50 kms$^{\rm -1}$), but we note that the construction of our model (Section 2), in particular the choice of $\lambda_{\rm s}$, accommodates the scenario in which the UDGs are underdeveloped HSB galaxies. Since log($\lambda_{\rm s}$)$\simeq$$\alpha$, our progenitor model at \z{2}{} can be assumed to have nominally developed a factor $\sim$10 more massive stellar disc if not quenched so early.

We presently find tentative support for an early accretion/quenching scenario with the deprojection (with an Abel integral identity) of the UDGs spatial distribution (Figure 7), which reveals a number density (as a function of radius) that can be reasonably fit with a NFW profile with c$_{\rm NFW}$ of 3, consistent with the satelite population at large \citep{lin04}. 

If ignoring the small sample size, this implies a relaxed distribution, in a cluster whose two-body relaxation time is of
order a Hubble time. However, the more centrally located dwarf satellites of Coma tend to be older \citep[in spite of a wide range of stellar ages depending on the chosen mass proxy;][]{smit12}. Figure 7 indicates the 47 known UDGs exhibit no clear trends among their properties as a function of cluter-centric radius, although VD15 note that the intracluster light may be obscuring some UDGs.

Our tidal diruption model is simple, insofar as the tidal harassment from other Coma satellites is not explicitly incorporated (and in which case our mass loss predictions may be lower limits). We have assumed at this stage that the UDGs, as low mass systems in a kinematically hot cluster \citep[$\sigma$$\simeq$1000 kms$^{\rm -1}$;][]{dane80}, are not significantly disrupted even on long timescales, as demonstrated in previous studies \citep[e.g.][]{torm98}. In general, dwarf galaxies are weakly influenced by dynamical friction in clusters (i.e. drag force $\propto$ satellite mass), with the associated orbit mostly avoiding regions of strong harassment \citep{smit10}. Diffuse systems such as the UDGs may remain quite vulnerable \citep{moor96}, but we have tentatively demonstrated here that the halo masses associated with these systems can provide some measure of stability. 

We cannot state however if this stability is sufficient in the event that the UDGs were first accreted as part of a group, as exhibited by simulated dwarf satellites \citep{delu12}. Our models have assumed a static host at the two epochs concerned (\z{0.023}{} and \z{2}), yet Coma-cluster analogues within cosmological simulations accrete $>$5 group-mass (10$^{\rm 13}$ M$_{\odot}$) halos within the last Gyr alone \citep{fakh10}. The dynamical state of major subgroups in Coma (e.g. NGC 4839) as pre- or post-merger is uncertain \citep{burn94, sand13}, but the (radial, anisotropic) kinematics of other faint dwarves tentatively suggest a significant recent infall from the field or as part of subgroups \citep{adam09}. 

Besides pre-processing in the more tidally disruptive environment of a group, tidal interactions between galaxies accreted as a group/filament are also quite feasible \citep{gned03}. This scenario is not compatible with our hypothesized high-$z$ quenching and low-$z$ late-type morphology of the UDGs, but is consistent with cosmological simulations which rarely find large intact LSBs in clusters \citep[e.g.][]{avil05}. 

To conclude, it is entirely possible that the observed UDGs are classical LSBs, recently accreted, and even on the verge of destruction. The recent deep imaging of the largest UDG evidence does not support its tidal disruption, although this may not be representative of the population \citep{vdok2015b}. We might also speculate if they are the cluster-bound gas-stripped remnants of another population, recently detected in the ALFALFA survey, constituting exceptionally gas-rich (for their stellar mass) dark galaxies, typically with high spin parameters \citep{huan12}. The kinematics of the UDGs would certainly aid in identifying their relationship with their hosts' dynamical history e.g. subgroup mergers \citep{vija15}. With this information, and the dynamical mass/metallicities obtainable in future long-exposure LSB-optimised observations, we look forward to refining our theoretical model.

\begin{figure}
\includegraphics[width=1.\columnwidth]{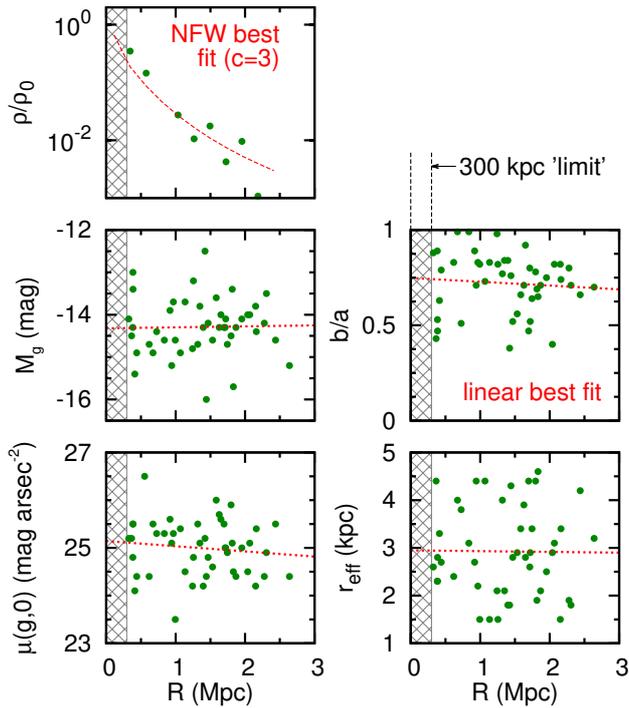}
\caption{(Top left) Deprojected number density of the 47 detected UDGs (using an Abel integral identity), normalised by the central density of a best-fit NFW profile with c$_{\rm NFW}$ of 3 (see text; red dashed line), as a function of cluster-centric radius; (middle left) $g$-band Magnitude, (middle right) \bia{}, (middle left) \muc{}, and (middle right) \reff{} as a function of projected cluster-centric radius for the 47 detected UDGs, with a linear-fit to each data set (red dashed line). }
\label{f1}
\end{figure}

\section*{Acknowledgements}

We thank the anonymous referee for their comments which improved this paper. We further thank Gerhardt Meurer for bringing the original UDG paper to our attention and useful comments. CY is supported by the Australian Postgraduate Award Scholarship. 

\bibliographystyle{mn2e}
\bibliography{bib}

\bsp
\label{lastpage}
\end{document}